\documentclass[aps,prb,twocolumn,floatfix]{revtex4}
\usepackage{bm}
\usepackage{epsf}
\usepackage{amssymb}
\usepackage{amsmath}
\usepackage{graphicx}
\usepackage{rotating}
\usepackage{epsfig}
\usepackage{psfrag}
\usepackage{amsmath}
\usepackage{hyperref}
\usepackage{subfigure}
\usepackage{braket}
\usepackage{tikz}
\usepackage{dsfont}
\usepackage{stmaryrd}
\usepackage{wasysym}
\usepackage{amsfonts}

\begin{document}

\title{Vison crystals, chiral and crystalline phases in the Yao-Lee model
}

\author{Muhammad Akram$^{1,2}$, Emilian Marius Nica$^{1,3}$, Yuan-Ming Lu$^4$, Onur Erten$^1$}
\affiliation{$^1$Department of Physics, Arizona State University, Tempe, AZ 85287, USA \\ $^2$Department of Physics, Balochistan University of Information Technology, Engineering and Management Sciences (BUITEMS), Quetta 87300, Pakistan\\ $^3$Department of Physics and Astronomy, Rice University, 6100 Main St, Houston 77005 TX, USA\\$^4$Department of Physics, The Ohio State University, Columbus OH 43210, USA}

\begin{abstract}
We study the phase diagram of the Yao-Lee model with Kitaev-type spin-orbital interactions in the presence of Dzyaloshinskii-Moriya interactions and external magnetic fields. Unlike the Kitaev model, the Yao-Lee model can still be solved exactly under these perturbations due to the enlarged local Hilbert space. Through a variational analysis, we obtain a rich ground state phase diagram that consists of a variety of vison crystals with periodic arrangements of background $\mathbb{Z}_2$ flux (i.e. visons). With an out-of-plane magnetic field, these phases have gapped bulk and chiral edge states, characterized by a Chern number $\nu$ and an associated chiral central charge $c_-=\nu/2$ of edge states. We also find 
helical Majorana edge states that are protected by magnetic mirror symmetry. For the bilayer systems, we find that interlayer coupling can also stabilize new topological phases. Our results spotlight the tunability and the accompanying rich physics in exactly-solvable spin-orbital generalizations of the Kitaev model.

\end{abstract}
\maketitle

\section{Introduction}
In the absence of magnetic order down to zero temperature, a quantum spin liquid (QSL) ground state displays unique features such as fractionalization and emergent gauge fields, due to 
underlying long-range quantum entanglement.~\cite{ Balents_Nature2010, Zhou_RMP2017, Wen_RMP2017, Knolle_AnnRevCondMatPhys2019, Broholm_Science2020}. The Kitaev model on a honeycomb lattice~\cite{Kitaev_AnnPhys2006} is an archetypal model for QSLs as the first exactly-solvable case with a QSL ground state with gapless as well as gapped phases with abelian and non-abelian anyons. In recent years, there has been significant progresses in identifying QSL materials with strong Kitaev interactions, such as iridates and $\alpha$-RuCl$_3$~\cite{HwanChun_NatPhys2015, Takagi_NatRevPhys2019,Kitagawa_Nat2018}. In spite of important progress, the confirmation of a QSL ground state remains under debate\cite{Yokoi_Science2021, Czajka_NatPhys2021}.

\begin{figure}[t]
\center
\includegraphics[width=0.5\textwidth]{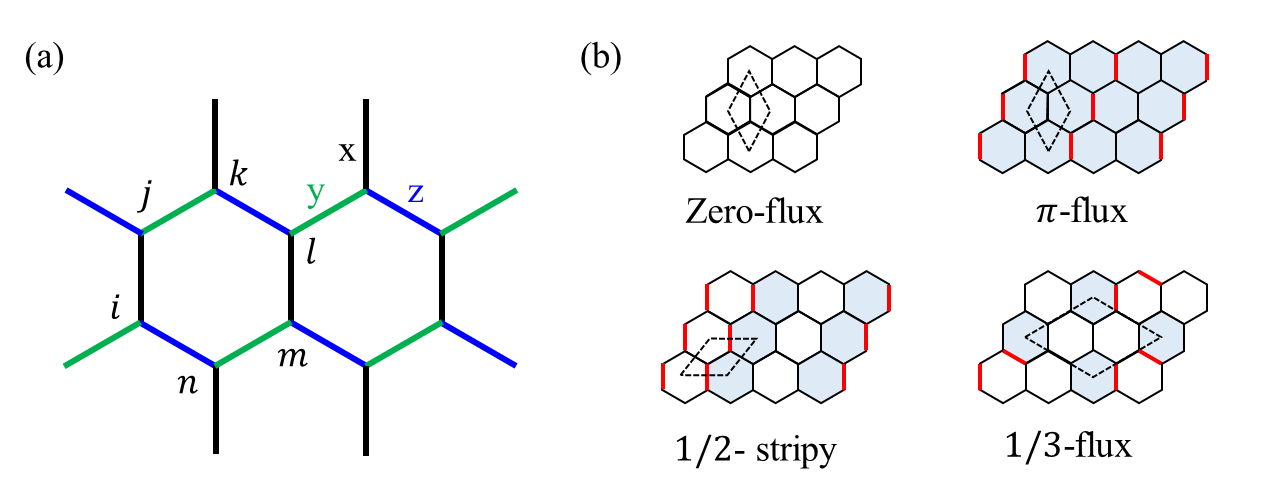} 
\caption{Illustration of the Yao-Lee model (a) honeycomb lattice with three types of bonds x, y and z (black, green and blue, respectively). (b) Examples of vison crystals. The red lines represent `flipped' bonds with $u_{ij}=-1$, which differ from the zero-flux gauge configuration where $u_{ij}=1$. The white and gray plaquettes represent zero ($W=1$) and $\pi$ ($W=-1$) fluxes, respectively. The dashed lines show the unit cells.}
\label{Fig:1}
\end{figure}

Although the Kitaev model has one of the most elegant solutions in quantum magnetism, the QSL state is quite fragile to perturbations. For instance, in model Hamiltonians proposed for candidate materials, the QSL phase occupies a small portion of the phase diagram\cite{Rau_PRL2014}. Moreover, most additional interactions destroy the integrability of the Kitaev model, making it very difficult to determine 
the ground state of the system. One remedy 
relies on extending Kitaev's original model to a larger family of exactly-solvable models that includes additional local, e.g. orbital, degrees of freedom (DOF). 
Such models feature the Kugel-Khomskii interaction\cite{Kugel_SovPhys1982}, which involves both spin and orbital DOF\cite{Ryu_PRB2009, Wang_PRB2009, Yao_PRL2009, Yao_PRL2011, PhysRevB.81.125134, Chulliparambil_PRB2020, Seifert_PRL2020, Carvalho_PRB2018, Natori_PRL2020, PhysRevB.106.125144, Jin:2021uua, PhysRevLett.129.177601, nica2023kitaev,Lu2022, PhysRevResearch.5.L022062}. These spin-orbital generalizations of the Kitaev model offer more flexibility in incorporating additional terms that preserve the exact solvability of the model. 

\begin{figure*}[t]
\center
\includegraphics[width=\textwidth]{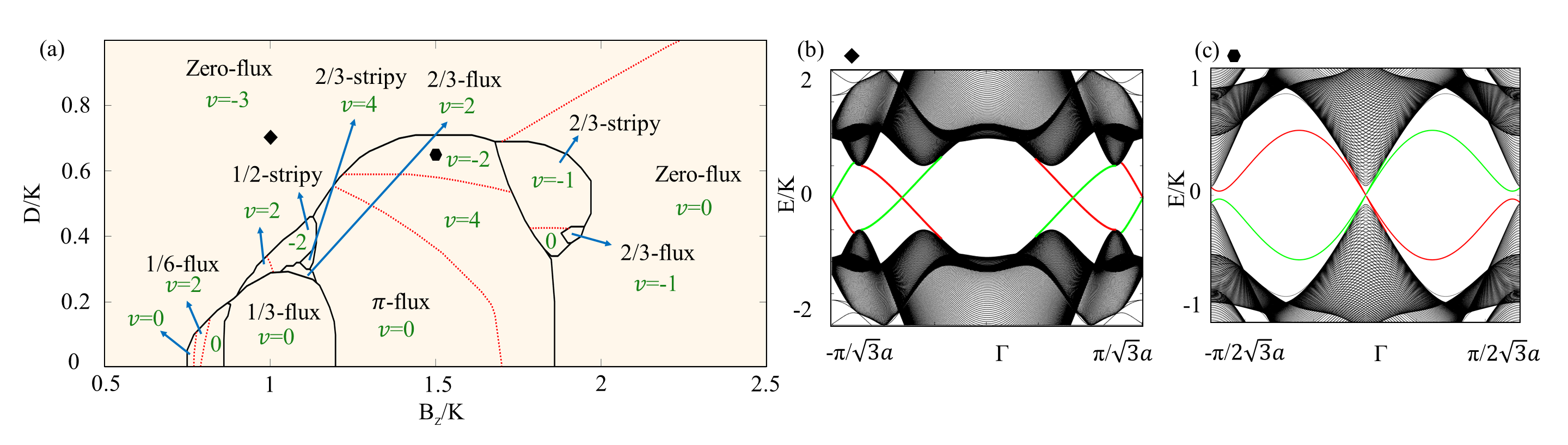} 
\caption{(a) Phase diagram as a function of out-of-plane magnetic field $B_z$ and DMI strength $D$ for a monolayer. For finite $D$ and $B_z$, all of the phases are gapped. The solid black lines represent first-order phase transitions between the vison crystal phases and the dashed red lines represent gap-closing topological phase transitions where the Chern number $\nu$ changes. (b) Chiral edge states in the zero-flux, $\nu=-3$ phase at $B_z/K =1$ and $D/K=0.7$. The red and green colors represent edge states on the two open boundaries, respectively. (c) Chiral edge states for the $\pi$-flux, $\nu=-2$ phase at $B_z/K=1.5$ and $D/K=0.65$.}
\label{Fig:2}
\end{figure*}

For this reason, we study the ground state phase diagram of a spin-orbital generalization of Kitaev model, first proposed by Yao and Lee~\cite{Yao_PRL2011} (YL), with an additional Dzyaloshinskii–Moriya interaction (DMI), an external magnetic field $\rm (B)$ in the monolayer system, and additional inter-layer coupling in the bilayer system via a variational analysis. DMI and the magnetic field are two important perturbations that can give rise to interesting magnetic phases, such as skyrmions in chiral magnets\cite{doi:10.1126/science.1166767,Yu2010-cn,PhysRevX.4.031045}. In the Kitaev model, these perturbations break integribility and therefore prohibit an exact solution. However, Majorana mean field theory studies suggest that a variety of chiral QSL phases can be stabilized\cite{Ralko_PRL2020}. On the contrary, for the YL model, these perturbations can be treated exactly due to the enlarged local Hilbert space. Our main results are as follows: (i) The phase diagram for finite $B_z$ and DMI displays a diverse set of gapped vison crystals. Moreover, these phases are further characterized by a Chern number, $\nu$ ranging from $0$ to $4$. (ii) We also find helical edge states that are protected by magnetic mirror symmetry for the zero-flux, $\nu=0$ phase. (iii) The in-plane magnetic field also gives rise to 
a rich phase diagram of vison crystals. However, these phases are mostly gapless, except for the 1/3-, 2/3- and 1/4-flux configurations with $\nu=0$. (iv) For the bilayer system, we perform self-consistent mean-field theory calculations for the inter-layer exchange and show that additional topological phases can be stabilized.

\section{Model}
We consider the YL\cite{Yao_PRL2011} model on a honeycomb lattice with spin-orbital DMI and an external magnetic field,
\begin{equation}
    H= H_{YL}+H_{DM}+H_{B}\\
\label{Eq:H}
\end{equation}
where
\begin{equation}
    H_{YL}= \sum_{\alpha {\rm -links}, \langle ij \rangle} K^{(\alpha)} \left(\tau_{i}^{(\alpha)} \tau_{j}^{(\alpha)}\right) 
\left( \pmb{\sigma}_{i} \cdot \pmb{\sigma}_{j} \right),
\end{equation}

\begin{equation}
    H_{DM}= D\sum_{\alpha {\rm -links}, \langle ij \rangle}  \left(\tau_{ i}^{(\alpha)} \tau_{j}^{(\alpha)}\right) 
\pmb{\hat{\delta}}_{(ij)}^{(\alpha)} \cdot \left( \pmb{\sigma}_{i} \times \pmb{\sigma}_{j} \right),
\end{equation}

\begin{equation}
    H_{B}= \sum_i \pmb{B}_{i} \cdot \pmb{\sigma}_{i}.
\end{equation}
Here, $K^{(\alpha)}$ is the nearest neighbor YL coupling constant, $D$ is 
the DMI coupling, and $\pmb{\hat{\delta}}^{(\alpha)}$ is the DMI vector for links $\alpha \in \{x,y,z\}$ as shown in Fig.~\ref{Fig:1}. The plaquette 
flux operator (Fig.~\ref{Fig:1}(a))
is defined as $W= \tau_i^x\tau_j^y\tau_k^z\tau_l^x\tau_m^y\tau_n^z\otimes \mathds{1}$. $W$ involves only orbital DOFs and commutes with the Hamiltonian, $[H,~W]=0$. As a result, the eigenstates of the Hamiltonian can be labeled by the $\pm 1$ eigenvalues of the plaquette operator. We consider a DMI vector $\pmb{\hat{\delta}}=\pmb{\hat{r}}_{ij}\times \pmb{\hat{z}}$ which arises from broken inversion symmetry on the surface\cite{Banerjee_NatPhys2013}. In addition, our DMI differs from the usual DMI 
due to the additional Kitaev-type coupling 
for the orbital DOF. This special form is critical as it commutes with the flux operator and consequently becomes bilinear 
in the itinerant Majorana fermions. In contrast, the usual DMI leads to quartic Majorana interactions which preclude a closed-form solution. We expect that our form of DMI would naturally occur in systems with spin and orbital DOFs which already include YL interactions, due to the broken inversion symmetry. 

It is possible to obtain an exact solution to eq.\eqref{Eq:H} via Majorana fermion representations for the spin and orbital DOF\cite{Yao_PRL2011}: $\sigma_{j}^{(\alpha)}=-i\epsilon^{\alpha \beta \gamma}c_{j}^{(\beta)}c_{j}^{(\gamma)}/2$, $\tau_{j}^{(\alpha)}=-i\epsilon^{\alpha \beta \gamma}d_{j}^{(\beta)}d_{j}^{(\gamma)}/2$ and $\sigma_i^\alpha \tau_j^\beta=ic_i^\alpha d_j^\beta$. However, these representations are redundant, and the physical states in each layer must be restricted to the eigenstates of $D_{i}=-i c_{i}^{(x)} c_{i}^{(y)} c_{i}^{(z)} d_{i}^{(x)} d_{i}^{(y)} d_{i}^{(z)}$ operators with eigenvalues $1$. Similar to Kitaev's original model, these constraints can be enforced through projection operators $P=\prod_i (1+D_{i})/2$. The monolayer Hamiltonian in the Majorana representation can be expressed as $H=  P \mathcal{H} P$ where $\mathcal{H}=\mathcal{H}_{YL}+\mathcal{H}_{DM}+\mathcal{H}_{B}$,
\begin{equation}
    \mathcal{H}_{YL}=\sum_{\alpha {\rm -links},\langle ij \rangle} K^{(\alpha)} u^{\alpha}_{ij}[ic_{i}^{(x)} c_{j}^{(x)}+i c_{i}^{(y)} c_{j}^{(y)} + ic_{i}^{(z)} c_{j}^{(z)}],
\end{equation}
\noindent \begin{align}
\mathcal{H}_{DM}= & D\sum_{\alpha {\rm -links},\langle ij \rangle} u^{\alpha}_{ij}[\hat{\delta}^{(\alpha)}_{(ij,x)}(ic_{i}^{(y)} c_{j}^{(z)}-i c_{i}^{(z)} c_{j}^{(y)})\notag \\
 & + \hat{\delta}^{(\alpha)}_{(ij,y)}(ic_{i}^{(z)} c_{j}^{(x)}-i c_{i}^{(x)} c_{j}^{(z)})],
\end{align}

\begin{equation}
    \mathcal{H}_{B}=\sum_{i} [i B_{x}c_{i}^{(y)} c_{i}^{(z)}+i B_{y}c_{i}^{(z)} c_{i}^{(x)}+i B_{z}c_{i}^{(x)} c_{i}^{(y)}].\\
\end{equation}
The bond operators $u^{(\alpha)}_{\nu, ij}=-id_{\nu,i}^{(\alpha)}d_{\nu,j}^{(\alpha)}$, commute with $\mathcal{H}$, and are therefore conserved with eigenvalues $\pm 1$. In the Majorana representation the plaquette operator is defined by the product of bond operators ($u_{ij}$) around the hexagonal plaquette.

In the absence of the external magnetic field and DMI, the ground state of the YL model lies in the zero-flux sector according to Lieb's theorem~\cite{Lieb_PRL1994}. Furthermore, the three Majorana fermions exhibit the same non-interacting spectra which is obtained by performing a Fourier transformation over half of the Brillouin zone. The spectra are gapless for $K_x+K_y>K_z$ and gapped otherwise. In our analysis, we consider the isotropic Kitaev interactions $K_x, K_y, K_z = K$. In this case, Majorana fermions have Dirac spectra, and are coupled to static $\mathds{Z}_2$ gauge fields. `Flipping' an odd number of the $u^{(\alpha)}_{ij}$ bonds around a plaquette induces a $\pi$-flux which is a vison excitation. This affects the kinetic energy of the Majorana fermions through a $\pi$ Berry phase. A periodic arrangement of the visons is called a vison crystal. In the original Kitaev model, vison crystals can be stabilized by additional interactions\cite{ PhysRevLett.122.257204,Zhang_PRL2019}. Vison crystals break translational symmetries and therefore have a local Ginzburg-Landau type order parameter and exhibit a finite-temperature phase transition. The order parameter of vison crystals is given by vison structure factor\cite{Zhang_PRL2019}
\noindent \begin{align}
    \rho(\textbf{k})=\frac{1}{L^2}\sum_{m,n}e^{i\textbf{k}.(\textbf{r}_m-\textbf{r}_n)}\langle W_m W_n \rangle
\end{align}
where $\textbf{r}_m$ is the position of $m^\textbf{th}$ hexagon, $W=\pm 1$, $L$ is the system size and the sum is over all hexagons.

In the presence of DMI and magnetic fields,  Lieb's theorem is not applicable and several vison crystal configurations can be stabilized. We consider 58 such vison crystal phases in our variational calculations\cite{Zhang_PRL2019, Chulliparambil_PRB2021,PhysRevLett.122.257204}. These are shown in appendix A figure~\ref{Fig:1:supp}.

For the bilayer system, we consider AA-stacked layers coupled via Heisenberg antiferromagnetic interactions $H_{\text{I}}=\sum_{i} J \pmb{\sigma}_{1i}\cdot \pmb{\sigma}_{2i}$. Here $(1,2)$ are layer indices, and $J$ is the nearest-neighbor inter-layer coupling constant. The inter-layer Hamiltonian in the Majorana representation is 
\noindent \begin{align}
    \mathcal{H}_{\text{I}}= J\sum_{i;\alpha \neq \beta} \left( c_{1i}^{(\alpha)}c_{2i}^{(\alpha)}c_{1i}^{(\beta)}c_{2i}^{(\beta)} \right).
    \label{eq:Hint}
\end{align}
 \begin{figure}[t]
\center
\includegraphics[width=0.5\textwidth]{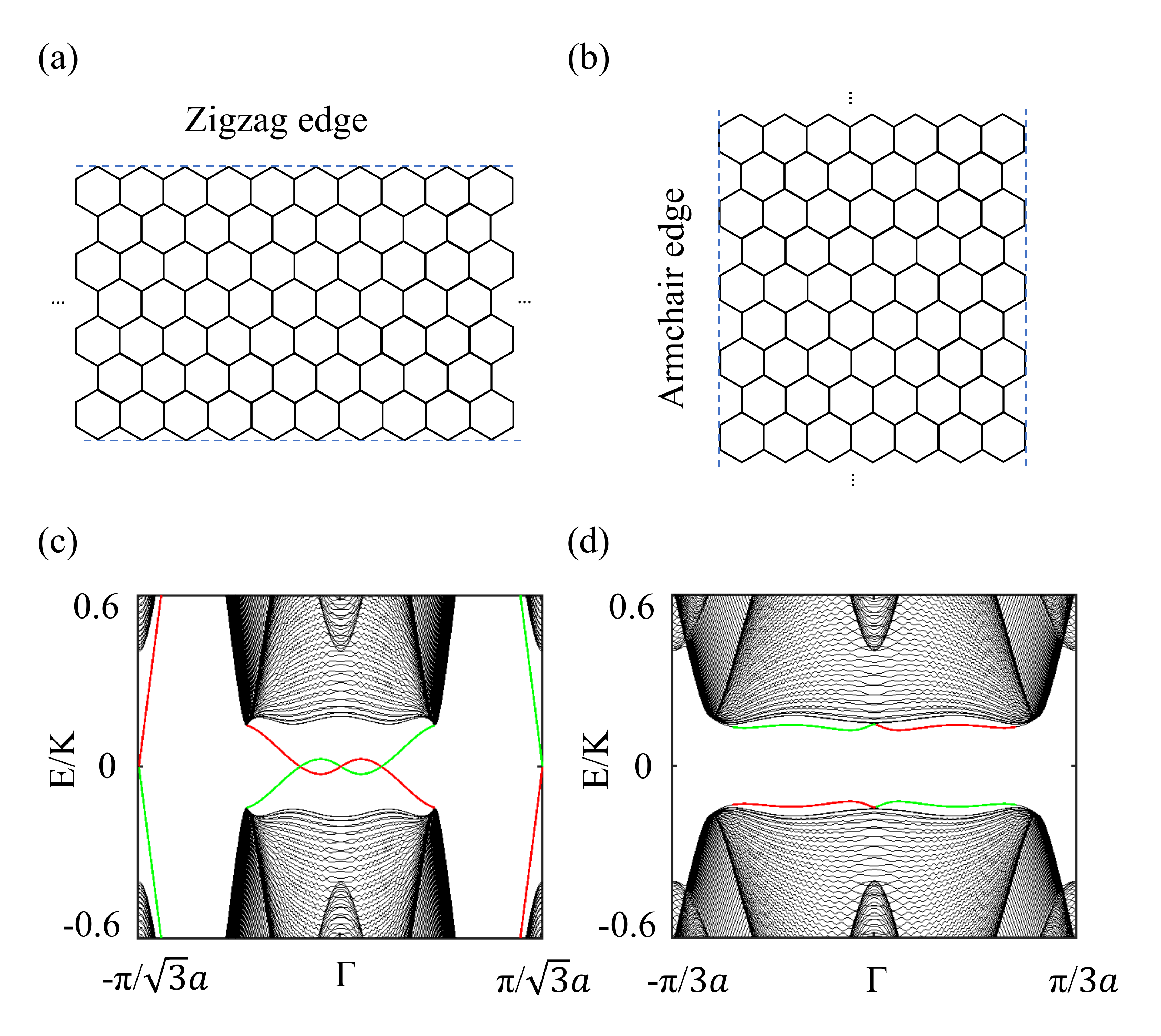}
\caption{Zigzag and armchair edges and their corresponding helical and gapped edge states for the zero-flux, $\nu=0$ phase. (a) A schematic picture of zigzag edge. (b) A schematic picture of armchair edge. (c) Helical edge states on zigzag open boundaries at $B_z/K=2.3$ and $D/K=0.8$. The red and green colors represent edge states on the two open boundaries, respectively. (d) Gapped states on armchair open boundaries at $B_z/K=2.3$ and $D/K=0.8$.}
\label{Fig:5}
\end{figure}
Although the inter-layer exchange interactions commute with the flux operators, they cannot be reduced to a bilinear form. Therefore, we decouple these interactions using mean field theory\cite{nica2023kitaev}. There are three possible mean field channels: $\langle ic_{1i}^{(\alpha)} c_{2i}^{(\alpha)}\rangle$, $\langle ic_{1i}^{(\alpha)} c_{2i}^{(\beta)}\rangle$ and  $\langle ic_{1i}^{(\alpha)} c_{1i}^{(\beta)}\rangle$ where $\alpha \neq \beta$. Here, we consider the Hartree channel $\langle \chi^{(\alpha)}_i \rangle = \langle ic_{1i}^{(\alpha)} c_{2i}^{(\alpha)}\rangle$. The second channel is related to the Hartree channel via an SO(3) rotation of the Majorana fermions in one of the layers. We ignore the magnetic channel, $\langle ic_{1i}^{(\alpha)} c_{1i}^{(\beta)}\rangle$, since we expect the interlayer interaction to lead to singlet formation, not long range magnetic order. 

\section{Results and Discussion}
Vison crystals in the YL model in the presence of magnetic field have been recently studied in Ref. \citenum{Chulliparambil_PRB2021}, which concluded that the ground state can have zero-, $\pi$- or 1/3-flux patterns as a function of a magnetic field. Here, we extend this analysis and consider additional DMI and inter-layer exchange interactions. The phase diagram for out-of-plane magnetic fields and DMI in the monolayer model is shown in Fig.~\ref{Fig:2}~(a). It exhibits seven distinct vison crystal phases, the most notable having zero-, $\pi$-, and 1/3-fluxes, respectively. It also includes 2/3-stripy, 2/3, 1/2-stripy and 1/6 flux configurations. Our phase diagram matches well with the phase diagram of Ref. \cite{Chulliparambil_PRB2021} for a wide range of $B_z$ in the absence of DMI. However, we obtain an additional 1/6-flux phase in our phase diagram that lies between zero and 1/3 fluxes. All of the phases are gapped for finite $B_z$ and $D$ and are further identified by the Chern number
\begin{eqnarray}
\nu=\frac{1}{\pi}\sum_{\alpha,\beta} \int_{BZ/2} d^2k~ \text{tr}F_{xy}^{\alpha \beta}(k).
\end{eqnarray}
of the Bogoliubov-de Gennes (BdG) Hamiltonian of the Majorana fermions. Here $F_{xy}^{\alpha \beta}=\partial_{k_x} A_y^{\alpha \beta}-\partial_{k_y}A_x^{\alpha \beta} +i([A_x,A_y])^{\alpha \beta}$ is the Berry curvature, $A^{\alpha \beta}=-i\langle \pmb{n}^\alpha(k)|\nabla_k|\pmb{n}^\beta(k)\rangle$ is the non-Abelian Berry connection and $\alpha$,$\beta$ represent filled bands defined over half Brillouin zone\cite{murakami20042,Chulliparambil_PRB2021}. The Chern number indicates the number of chiral edge modes as shown in Fig. ~\ref{Fig:2}~(b)-(c). 
There are topological gap-closing transitions within the vison crystal phases, represented by the red dashed lines in Fig. ~\ref{Fig:2}(a) where the Chern number changes. For instance, in the zero-flux sector, there are two topological phases with $\nu=-3$ and $\nu=0$ that are separated by a topological phase transition. Similarly there are three topological phases for $\nu=-2,0$, and $4$ in $\pi$-flux sector. Moreover, the 2/3-stripy, 2/3, 1/2-stripy and 1/6 phases also exhibit topological phases, while the 1/3-flux phase is trivial with $\nu=0$. 

\begin{figure}[t]
\center
\includegraphics[width=0.5\textwidth]{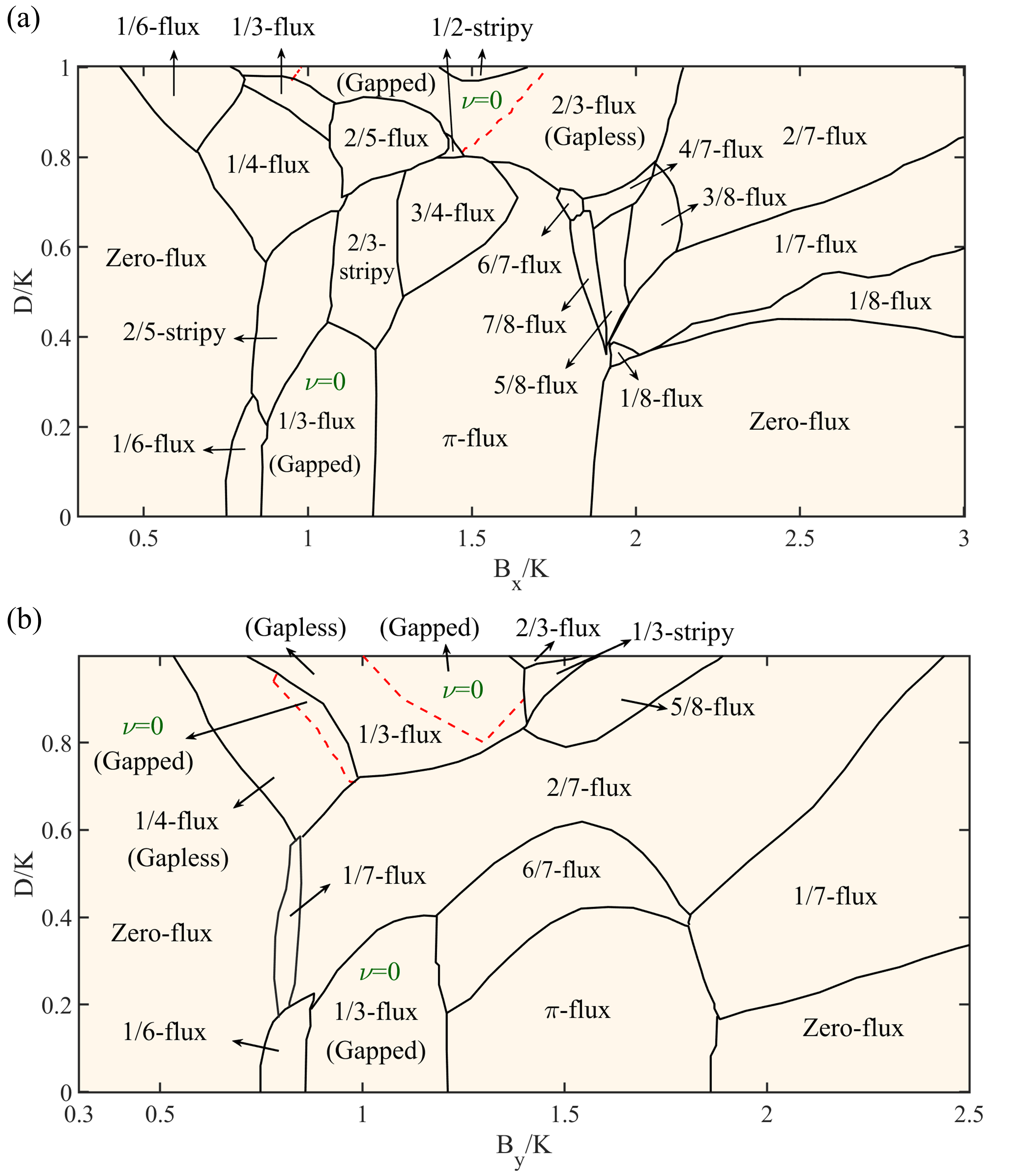}
\caption{(a) Phase diagram for in-plane magnetic field $B_x$ and DMI $D$ on the monolayer. The solid black lines represent first-order phase transitions and the dashed red lines represent topological phase transitions. (b) Phase diagram of the in-plane magnetic field $B_y$ and DMI $D$ on the monolayer. Unless explicitly indicated, the phases are gapless, in contrast to the case with out-of-plane fields.}
\label{Fig:3}
\end{figure}
The zero-flux, $\nu=0$ phase possesses helical edge states for the zigzag open boundary whereas they are gapped on the armchair open boundary as shown in Fig.~\ref{Fig:5}. These helical edge states are topological, protected by the following magnetic mirror symmetry $M'_x$:
\begin{eqnarray}
M'_x=M_x e^{i\frac{\pi}{2}\sum_i\sigma_i^x+i\frac{\pi}{4}\sum_i\tau_i^z}\mathcal{K}(-1)^s
\end{eqnarray}
where $M_x$ is spatial reflection about the x-axis, $\mathcal{K}$ represents complex conjugation, and $s=0,1$ for A and B sublattices, respectively. $M_x$ is preserved for zigzag open boundary whereas it is broken for armchair case. In momentum space, it has the form $M'_x=R_x(\pi)\otimes \mu_z\mathcal{K}$ where $R_x(\pi)$ is a $180^o$ rotation about the x-axis which acts on the Majorana DOF and $\mu_z$ is a Pauli matrix in the sublattice indices. The Hamiltonian under this symmetry transforms as:
\begin{eqnarray}
M'_x\mathcal{H}(k_x,k_y)(M'_x)^{-1}=\mathcal{H}(k_x,-k_y).
\end{eqnarray}
The helical edge states have zero energy crossings at high symmetry points $k_x=
\{0, \pi/\sqrt{3}a \}$ as shown in Fig.~\ref{Fig:5}(a). For $k_x=0$ and $k_x=\pi/\sqrt{3}a$, the BdG Hamiltonian reduces to two 1D Hamiltonians, $H(k_y)_{k_x=0,\pi}$ which have effective time reversal ($M'$), particle-hole ($\mathcal{K}$) and chiral ($\mathcal{C}=M'\mathcal{K}$) symmetries. According to Altland–Zirnbauer's 10-fold way classification\cite{Altland_PRB1997,Schnyder2008,Kitaev2009}, they belong to symmetry class BDI with a 1D topological invariant $\mathbb{Z}$ which can be calculated by the winding number\cite{PhysRevLett.89.077002,PhysRevB.78.195424,PhysRevLett.109.150408}
\begin{eqnarray}
w=\frac{-i}{\pi} \int_{0}^{\pi} \frac{dz(k)}{z(k)},
\end{eqnarray}
\begin{equation}
z=\frac{\rm{det}(F)}{F},
\end{equation}
\begin{align}
U\mathcal{H}(k)U^{\dagger}=\begin{pmatrix} 0 & F(k) \\ F(-k)^T & 0 \end{pmatrix}
\end{align}
where $U$ is the diagonal representation of chiral symmetry $\mathcal{C}$. We obtained $w=1$ for both $k_x=0$ and $\pi/\sqrt{3}a$ lines. Therefore, we have shown that the helical edge states on the zigzag boundary are protected by the magnetic mirror symmetry. On the other hand, for the armchair open boundary, the Hamiltonian preserves another anti-unitary mirror symmetry $M_y'=R_y(\pi)\otimes \mu_x\mathcal{K}$, particle-hole symmetry $\mathcal{K}$ and chiral symmetry $\mathcal{C}^\prime=R_y(\pi)\otimes \mu_x$ on the high-symmetry lines $k_y=0$ and $\pi/3a$. In this case, we get $w=0$ for both $k_y=0$ and $\pi/3a$, in agreement with the gapped edge states shown in Fig.~\ref{Fig:5}(b).

Next, we discuss the interplay between an in-plane magnetic field and DMI for the monolayer model as shown in Fig.~\ref{Fig:3}. Similar to the case of out-of-plane magnetic field, here we also obtain a rich phase diagram of vison crystals for both in-plane magnetic fields $B_x$ and $B_y$. However, the phase diagrams are gapless except for the 1/3-flux phase for  $B_x$ and 1/3- and 1/4-flux phases for $B_y$. Similar to the out-of-plane magnetic field cases, the dominant phases have zero, $\pi$, and 1/3 fluxes, but there are also 1/7, 2/7, 6/7, 2/3 and 2/3-stripy phases. In contrast to the out-of-plane magnetic fields, there are also additional phases present, namely the 1/3-stripy, 1/4, 3/4, 2/5-stripy, 1/8, 3/8, 5/8 and 7/8 flux phases.
\begin{figure}[t]
\center
\includegraphics[width=0.5\textwidth]{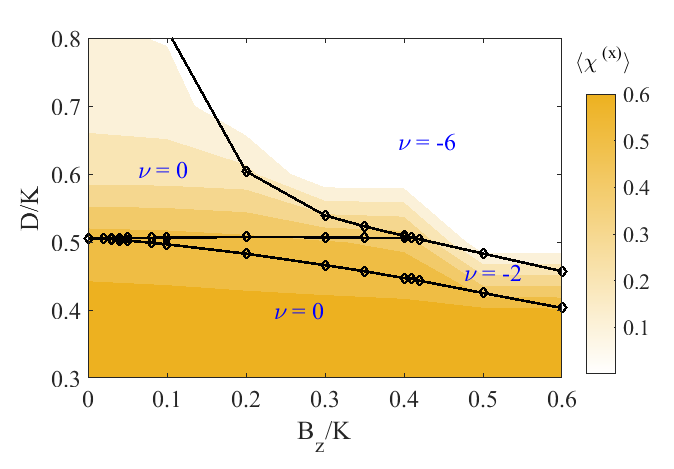}
\caption{Phase diagram for out-of-plane magnetic field $B_z$ and DMI $D$ for a bilayer model with zero flux at inter-layer coupling $J/K=1$. The color map indicates the value of the mean-field effective hybridization $\langle \chi^{(x)} \rangle$.}
\label{Fig:4}
\end{figure}

Lastly, we discuss a bilayer system with inter-layer couplings in an AA stacking configuration. We restrict our calculations to small $D$ and $B$ such that the zero-flux state remains the ground state. We follow a mean-field procedure for the inter-layer exchange (eq.\ref{eq:Hint}), first introduced in Ref. \citenum{nica2023kitaev}, and decouple the interaction in the Hartree channel $\langle \chi^{(\alpha)}_i \rangle = \langle ic_{1i}^{(\alpha)} c_{2i}^{(\alpha)}\rangle$. We consider opposite-sign $\langle \chi \rangle$ on the two sublattices, $\langle \chi_{AA} \rangle= -\langle \chi_{BB} \rangle$, which gaps the Majorana spectrum, thus lowering the ground state energy when compared to the case where $\langle \chi^{(\alpha)}_i \rangle$ has the same sign on both sublattices. Fig.~\ref{Fig:4} shows the phase diagram and self-consistent solutions $\langle \chi^{(x)} \rangle$ as functions of $B_z$ and DMI for $J/K=1$. Both $B_z$ and $D$ suppress $\langle \chi^{(x)}\rangle$ since they gap the Dirac spectrum which in return suppress the instability. A similar behaviour is obtained for $\langle \chi^{(y)} \rangle$ and $\langle \chi^{(z)} \rangle$. For large $B_z$ and $D$, a chiral topological phase with $\nu=-6$ emerges, corresponding to two uncoupled monolayers each with $\nu=-3$. On the other hand, for small $D$ and finite $B_{z}$, $\chi$ can lead to trivially gapped phases with $\nu=0$. This phase is adiabatically connected to the trivial gapped phase at $B_z=0$ and $D=0$\cite{nica2023kitaev}. As we increase $D$, $\chi$ decreases and we obtain a new topological phase with $\nu=-2$, where only one of the chiral edge modes from each layer survives. 
Having discussed our results in detail, we now comment on their general features. As already mentioned, the phase diagrams for out-of- and in-plane magnetic fields differ substantially, while showing similar complexity. The difference is due to the relative orientation of the fields and DMI vector, which naturally imposes a preferred axis. Their similar complexity, due to the presence of both spin and orbital DOF, further demonstrates the remarkable tunability of these models. However, the most striking results occur for the out-of-plane fields. Here, arbitrarily small DMI and applied field always lead to gapped phases. Furthermore, a topologically non-trivial phase ($\nu=-3$) zero-flux phase persists for a significant range of small fields and any non-zero DMI coupling. This is to be contrasted to the phase diagram of the YL model without DMI, where the topologically non-trivial $\pi$-flux phase ($\nu=4$) requires fine-tuning. In the case of the bilayer, the stability of the gapped trivial phase ($\nu=0$) can also be understood from the nature of the ground-state, which includes inter-layer spin-singlets on overlapping sites\cite{nica2023kitaev}. Nevertheless, a topologically non-trivial phase with $\nu=-2$ emerges here as well for the coupled layers.
There are several experimental signatures of the vison crystals with different Chern numbers\cite{Zhang_PRL2019}. For the out of plane magnetic field, the vison crystals are gapped and the heat capacity acts like $C \propto e^{-\Delta/T}$ where $\Delta$ is the smaller of the vison or Majorana fermion gap. The energy gap of the excitation spectrum can be detected from the temperature dependence of the specific heat. For the in-plane magnetic field, most of the vison crystals are gapless and contain both Dirac dispersion or Fermi surfaces with specific heat $C \propto T^2$ and $C \propto T$ respectively. 
Moreover, the distinct Majorana spectra can be distinguished based on their characteristic low energy signatures in spectroscopic probes, such as resonant inelastic x-ray scattering.\cite{PhysRevLett.117.127203,PhysRevLett.119.097202} In addition, the  vison crystals spontaneously break the translational invariance, their presence implies a finite-temperature phase transition into a high-temperature symmetric phase. Furthermore, vison crystals are expected to exhibit clear features of translation symmetry-breaking in neutron scattering experiments\cite{Zhang_PRL2019}. Moreover, a finite Chern number for a gapped bulk ground state can lead to a quantized thermal Hall conductance $K_{xy}=\nu\frac{\pi K_B^2T}{12\hbar}$ due to the chiral edge states\cite{Kane1997,Kitaev_AnnPhys2006}. YL models, as well as other spin-orbital extensions of the Kitaev model, generally require a quartet ground state at atomic level and strongly bond-dependent interactions. These conditions may be realized in strongly spin-orbit-coupled Mott insulators with orbital degeneracy or in materials with $J=3/2$ local moments\cite{Natori_PRB2019, Seifert_PRL2020, Xu_PRL2020, Stavropoulos_PRL2019}. For instance, an enhanced SU(4) symmetry has been proposed for $\alpha$-ZrCl$_3$~\cite{Yamada_PRL2018}. However, to date there are no predictions for material realizations of YL model.

\section{Conclusion}
Based on variational calculations, we have shown that a variety of vison crystal phases can be stabilized by an external magnetic field and DMI in the YL model. The vison crystals are gapped for an out-of-plane magnetic field and mostly gapless for an in-plane magnetic field. Gapped vison crystals include topological phases with both chiral and helical edge states. The latter are protected by a magnetic mirror symmetry and are present only on the zigzag open boundary. Furthermore, for the bilayer model, we have shown that additional topological phases can be stabilized by tunning the inter-layer coupling. An important future study involves the topological defects of these symmetry-breaking vison crystals, including domain walls, dislocations, and disclinations that can host gapless modes or non-Abelian anyons\cite{Hou2007,Wang2022}.

\section{Acknowledgements}
We thank Natalia Perkins, Rebecca Flint and Ribhu Kaul for fruitful discussions. OE acknowledge
support from NSF Award No. DMR 2234352. MA is supported by Fulbright Scholarship. YML is suppored by NSF under Award No. DMR2011876. EMN acknowledges support by NSF under Grant No.
DMR-2220603.

\appendix
\section{Variational Analysis}
\begin{figure*}
\center
\includegraphics[width=\textwidth]{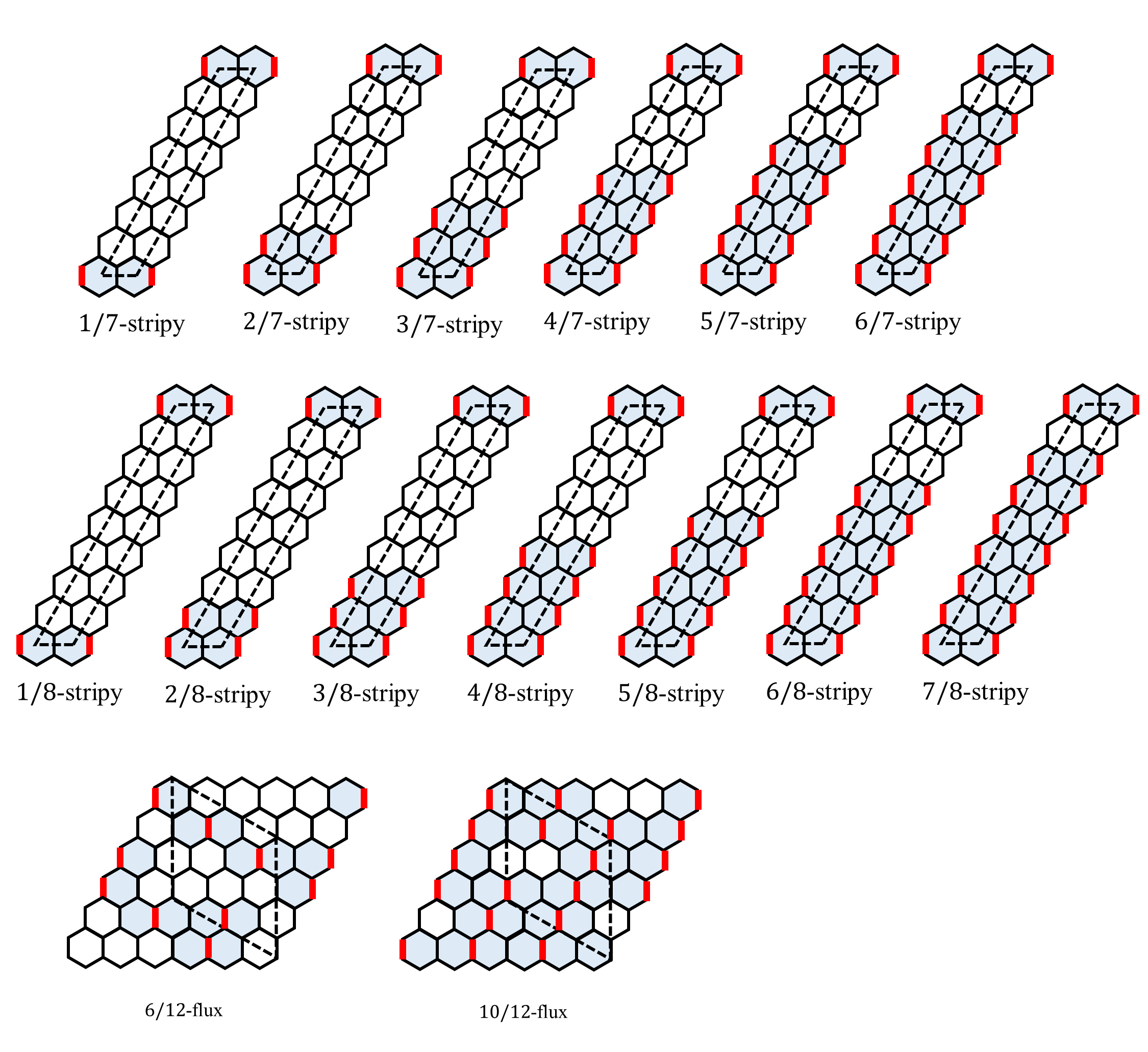} 
\caption{Vison crystals considered in variational analysis. The red bonds represent ``flipped" bonds with $u_{ij}=-1$, which differ from the flux-free gauge configuration where $u_{ij}=1$. The white and gray plaquettes represent zero ($W=1$) and $\pi$ ($W=-1$) fluxes. The dashed lines represent the unit cells.}
\label{Fig:1:supp}
\end{figure*}

In our variational analysis we consider 58 vison crystals as shown in Fig.~\ref{Fig:1:supp}. We find the energy of each vison crystal for each point of the phase diagram and then compare their energies to construct the phase diagrams.
\begin{figure*}
\center
\includegraphics[width=\textwidth]{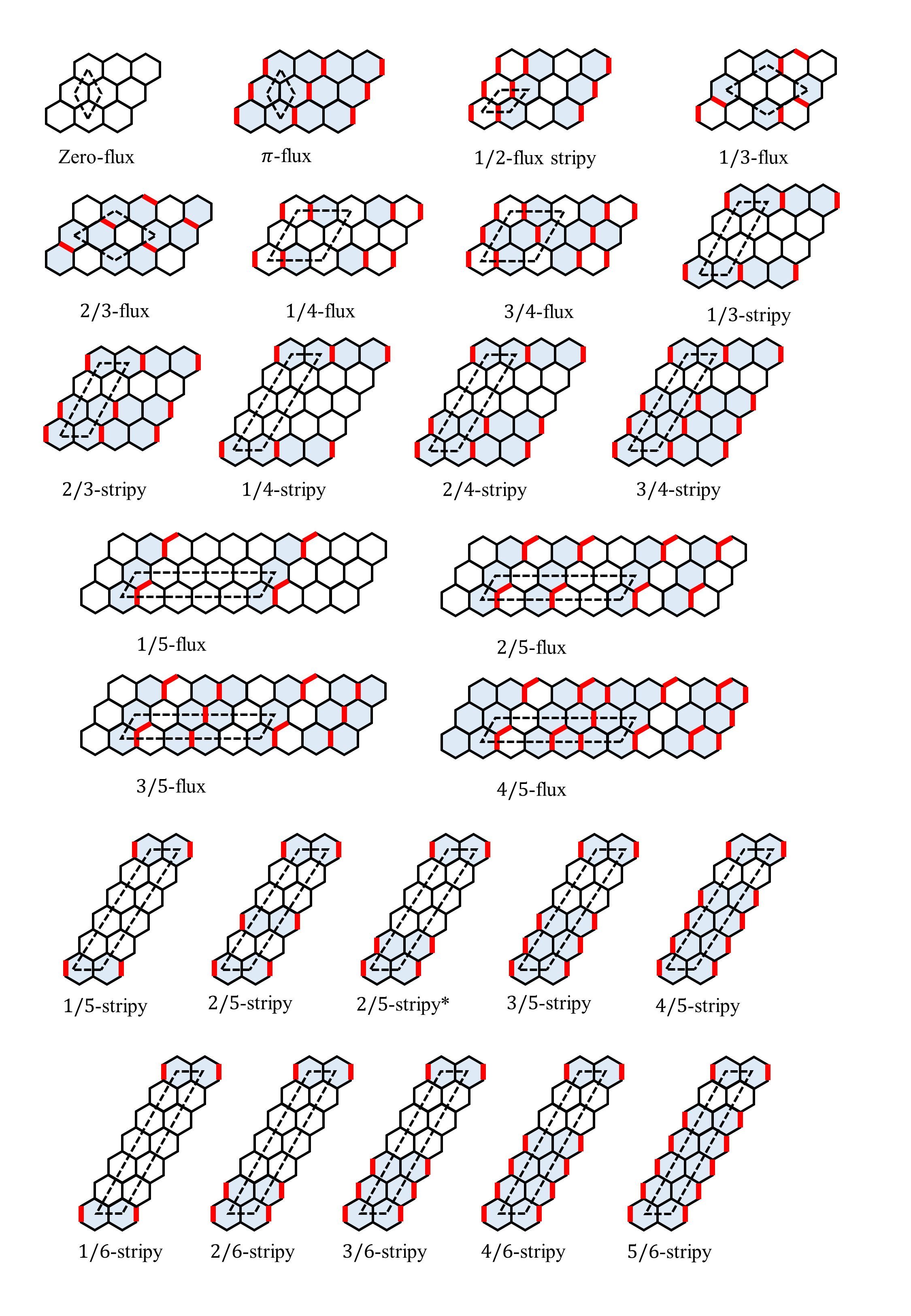} 

\end{figure*}

\begin{figure*}
\center
\includegraphics[width=\textwidth]{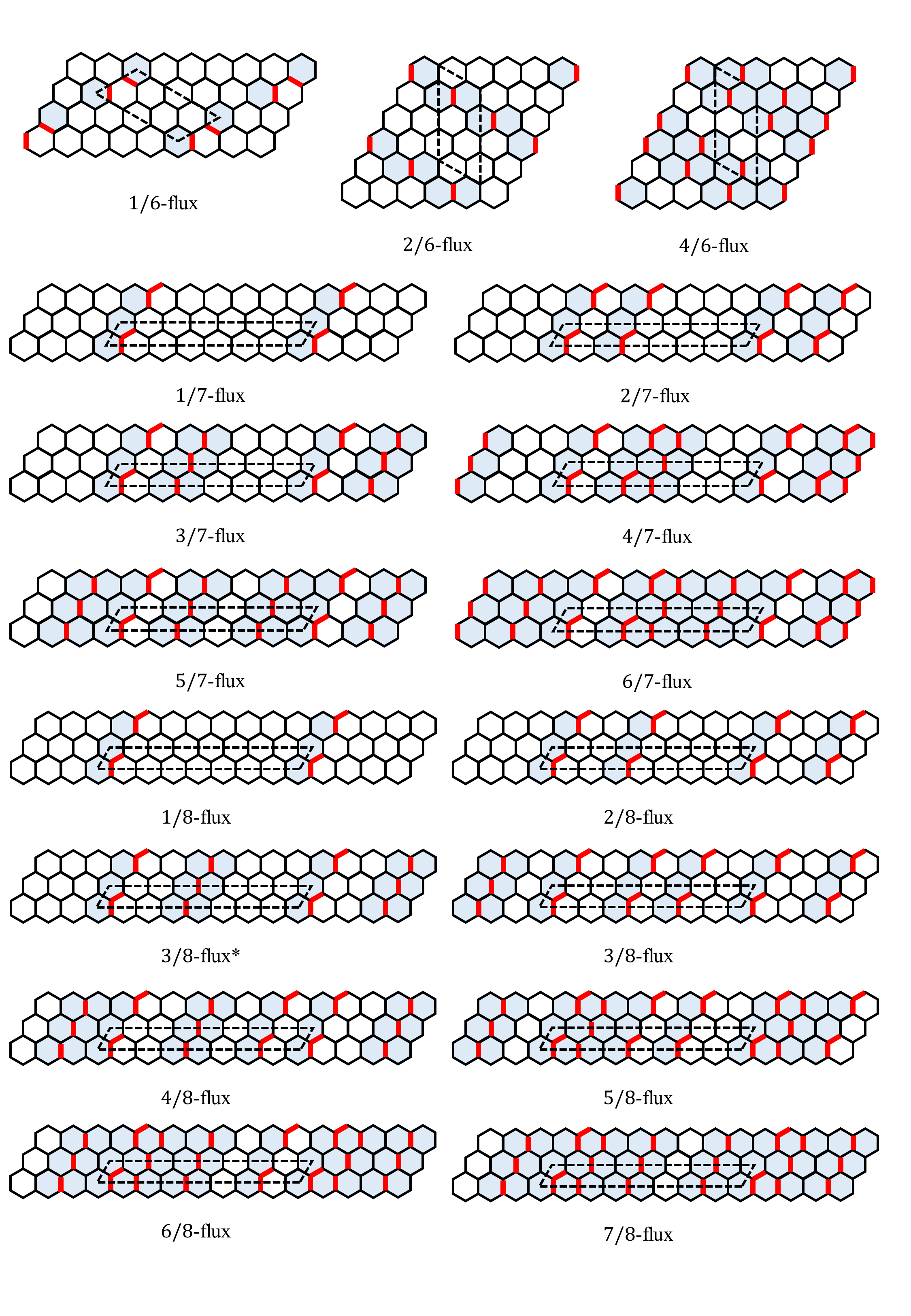} 
\end{figure*}

%

\end{document}